\documentclass[%
 reprint,
 amsmath,amssymb,
 aps,
]{revtex4-2}

\usepackage{graphicx}
\usepackage{dcolumn}
\usepackage{bm}
\usepackage{natbib}
\usepackage{multirow}
\usepackage{quantikz}
\usepackage{braket}
\usepackage{hyperref}
\usepackage{siunitx}
\usepackage{subcaption}
\usepackage{diagbox}
\usepackage{xspace}
\definecolor{color1}{HTML}{648FFF}
\definecolor{color2}{HTML}{785EF0}
\definecolor{color3}{HTML}{DC267F}
\definecolor{color4}{HTML}{FE6100}
\definecolor{color5}{HTML}{FFB000}
\definecolor{color6}{HTML}{882255}
\definecolor{color7}{HTML}{0C7BDC}
\definecolor{white}{HTML}{FFFFFF}

\newcommand{\qqbar}{\ensuremath{q\overline{q^\prime}}\xspace}

\newcommand{\pt}{$p_\mathrm{T}$}

\begin{document}

\preprint{APS/123-QED}

\title{One particle - one qubit: Particle physics data encoding for quantum machine learning}

\author{Aritra Bal}
\email{aritra.bal@kit.edu}
\author{Markus Klute}
\affiliation{%
 Institute of Experimental Particle Physics (ETP), Karlsruhe Institute of Technology, Karlsruhe, DE
}%

\author{Benedikt Maier}
\email{benedikt.maier@cern.ch}
\author{Melik Oughton}
\author{Eric Pezone}
\affiliation{%
 Imperial-X and Department of Physics, Imperial College Of Science, Technology And Medicine, London, UK
}%

\author{Michael Spannowsky}
\email{michael.spannowsky@durham.ac.uk}
\affiliation{%
 Institute for Particle Physics Phenomenology (IPPP), Durham University, Durham, UK
}%

             
\begin{abstract}
We introduce 1P1Q, a novel quantum data encoding scheme for high-energy physics (HEP), where each particle is assigned to an individual qubit, enabling direct representation of collision events on quantum circuits without classical compression. We demonstrate the effectiveness of 1P1Q in quantum machine learning (QML) through two applications: a Quantum Autoencoder (QAE) for unsupervised anomaly detection and a Variational Quantum Circuit (VQC) for supervised classification of top quark jets. Our results show that the QAE successfully distinguishes signal jets from background QCD jets, achieving superior performance compared to a classical autoencoder while utilizing significantly fewer trainable parameters. Similarly, the VQC achieves competitive classification performance, approaching state-of-the-art classical models despite its minimal computational complexity. Furthermore, we validate the QAE on real experimental data from the CMS detector, establishing the robustness of quantum algorithms in practical HEP applications. These results demonstrate that 1P1Q provides an effective and scalable quantum encoding strategy, offering new opportunities for applying quantum computing algorithms in collider data analysis.
\end{abstract}

\maketitle
\onecolumngrid
\vspace{-3em}
\begin{center}
\small\textit{Published in:} Physical Review D \textbf{112 (2025)}, 076004.
\end{center}
\twocolumngrid

\paragraph*{\textbf{\label{sec:level1}Introduction.}}

The unprecedented collision energies achieved at latest- and next-generation colliders, like the Large Hadron Collider (LHC) and the Future Circular Collider (FCC), produce vast amounts of particle-level data, challenging the limits of conventional data analysis techniques in high energy physics (HEP). With the increasing complexity of these datasets, novel approaches that leverage cutting-edge computational paradigms have become indispensable. Machine Learning (ML) has established itself as an indispensable tool for analysing HEP data in recent years, leading to significant advancements in tasks such as event classification, anomaly detection, and parameter estimation. An emerging frontier in this field is to extend these techniques to the quantum domain, leveraging Quantum Machine Learning (QML) to enhance data analysis capabilities. Quantum computing, with its ability to exploit superposition, entanglement, and interference, offers a promising framework to address some of the most intricate challenges in HEP~\cite{Guan:2020bdl,DiMeglio:2023nsa,Brown:2023llg}. 

Any quantum machine learning algorithm consists of three key components: (1) the data encoding, which maps classical data onto quantum states; (2) the quantum model, typically implemented through quantum circuits and quantum operations such as entanglement between qubits; and (3) the loss function, whose optimization is crucial for training the quantum model. While substantial efforts have been devoted to designing quantum models and optimizing loss functions \cite{Abel:2022lqr,Abel:2021fpn,Delgado:2022aty,Alvi:2022fkk,Araz:2022haf,Hammad:2023wme,Schuhmacher:2023pro,Chen:2024rna,Yang_2024bqw,Scott_2024txs}, the choice of data encoding remains a critical yet underexplored aspect of QML for HEP applications. In this work, we propose a new encoding scheme tailored to HEP data, which we call 1P1Q (1 Particle - 1 Qubit). This encoding strategy assigns a separate qubit to each particle, enabling an effective representation of collider events within quantum circuits without prior data compression. Establishing a direct correspondence between particles and qubits is warranted to advance the usefulness of quantum computing for HEP in particular upon entering an era in which crucial LHC measurements reveal the quantum mechanical nature of the imprints left behind by the collisions, such as the recently observed entanglement in top quark pair production~\cite{ATLAS:2023fsd,CMS:2024pts}.

We demonstrate the effectiveness of 1P1Q-based QML models in processing HEP data by employing two distinct approaches: a Quantum Autoencoder (QAE) \cite{Ngairangbam:2021yma} for unsupervised, unlabelled learning and a Variational Quantum Circuit (VQC) \cite{Blance:2020nhl} for classification tasks. The QAE compresses quantum states by learning a lower-dimensional latent representation, making it well-suited for anomaly detection in collider physics, while the VQC employs a parametrized quantum circuit to discriminate between different class categories based on learned quantum features. Both methods fully exploit the kinematic information of particles encoded via the 1P1Q scheme, using the quantum state representation to retain and process intricate correlations between particles. Such information is at risk of being lost in existing approaches~\cite{Blance:2020ktp,Chen:2021ouz,Gianelle:2022unu,Peixoto:2022zzk,Belis:2023atb,Duffy:2024zog,Hammad:2024dsn,Belis:2024guf}, which first encode the information of the collider events into an abstract, typically compressed, latent representation using classical machine learning algorithms or domain-inspired high-level features. Instead, 1P1Q guarantees more immediate access to the raw information of collider events and allows for the direct exploitation of their content using quantum variational circuits. 1P1Q therefore also provides a natural way of extending the input space as quantum computers and their simulators can accommodate more and more qubits. 

We apply the QML models to one of the most well-established use cases of classical machine learning in HEP: the discrimination of hadronically decaying resonances from the overwhelming background of QCD jets. Identifying such resonances, including those from top quarks, Higgs bosons, or hypothetical new physics states, is crucial for advancing our understanding of fundamental interactions. We find that for this task, VQC and QAE models acting on 1P1Q-encoded particle information result in highly performant QML algorithms, equal to or even better than comparable classical counterparts in the case of the QAE.

For the first time, our study explores QML in HEP using actual experimental data recorded by the CMS detector~\cite{CMS:2008} in 2016. This real-world application provides a critical test of the robustness and feasibility of QML strategies in general, and 1P1Q in particular, beyond simulated datasets. By demonstrating that QML models can be trained on and extract meaningful physics from real collision events, we show that 1P1Q can become a modern, lossless collider data encoding framework as quantum computing hardware advances.

In the following, we present the formulation of the 1P1Q encoding, explore its theoretical underpinnings, and discuss its potential applications to jet physics at the LHC and future colliders with the examples of anomaly detection and a supervised classifier. We demonstrate how this encoding captures the essential kinematic features of jets and showcase its utility in leveraging quantum algorithms to analyze jet substructure, and compare to state-of-the-art classical machine learning algorithms. 

\paragraph*{\textbf{1P1Q - Particle Encoding.}}

In collider measurements, reconstructed particles are kinematically fully described by three key parameters: the transverse momentum  $p_\mathrm{T}=\sqrt{p_x^2 + p_y^2}$, where  $p_x$  and  $p_y$  are the momentum components in the transverse detector plane, the pseudorapidity  $\eta$, and the azimuthal angle $\phi$. The pseudorapidity  $\eta$ is related to the polar angle $\nu$ of the particle’s trajectory by $\eta = -\ln \tan(\nu/2)$, which provides an approximation of the rapidity in high-energy regimes. Finally, the azimuthal angle $\phi$  describes the particle’s direction in the transverse plane and is measured relative to a chosen reference axis. Together, these three quantities specify the particle’s momentum.

The 1P1Q method directly encodes the kinematics of a particle on a qubit. The pseudorapidity $\eta$ and the azimuthal angle $\phi$ of the particle, modulated by the transverse momentum \pt{} normalized to the \pt{} of the jet, are used as spherical coordinates on the Bloch sphere to orient the qubit, enabling a compact and information-rich representation of each particle as a quantum state that is not dependent on the energy scale of the jet, thus facilitating a straightforward transition to the quantum domain. These feature encodings, represented by rotation angles about the $Y$ and $X$ axes respectively, are then additionally scaled by a factor  $f=f(w)$ constrained to lie between $[1,2\pi+1]$, where $w$ is a trainable parameter.  This ensures the particles can spread out across the Bloch sphere instead of clustering too closely around its North Pole. 
The encoding can then be summarized as:


\begin{align}
f \cdot \dfrac{p_\mathrm{T}}{p_\mathrm{T,jet}}\cdot(\eta-\eta_{\text{jet}})\quad&\to\quad\theta \label{eq:enc1} \\
f \cdot \dfrac{p_\mathrm{T}}{p_\mathrm{T,jet}}\cdot(\phi-\phi_{\text{jet}})\quad&\to\quad\varphi  \label{eq:enc2} \\
(p_\mathrm{T},\eta,\phi)\quad&\to\quad\ket{\psi}= R_X(\varphi) R_Y(\theta) \ket{0} \nonumber \\
&~~~~= \alpha(\theta,\varphi)\ket{0}+\beta(\theta,\varphi)\ket{0} \\
f \quad&\to\quad 1+\dfrac{2\pi}{1+e^{-w}} \label{eq:scale}
\end{align}

\begin{figure*}[ht]
\includegraphics[width=0.9\textwidth]{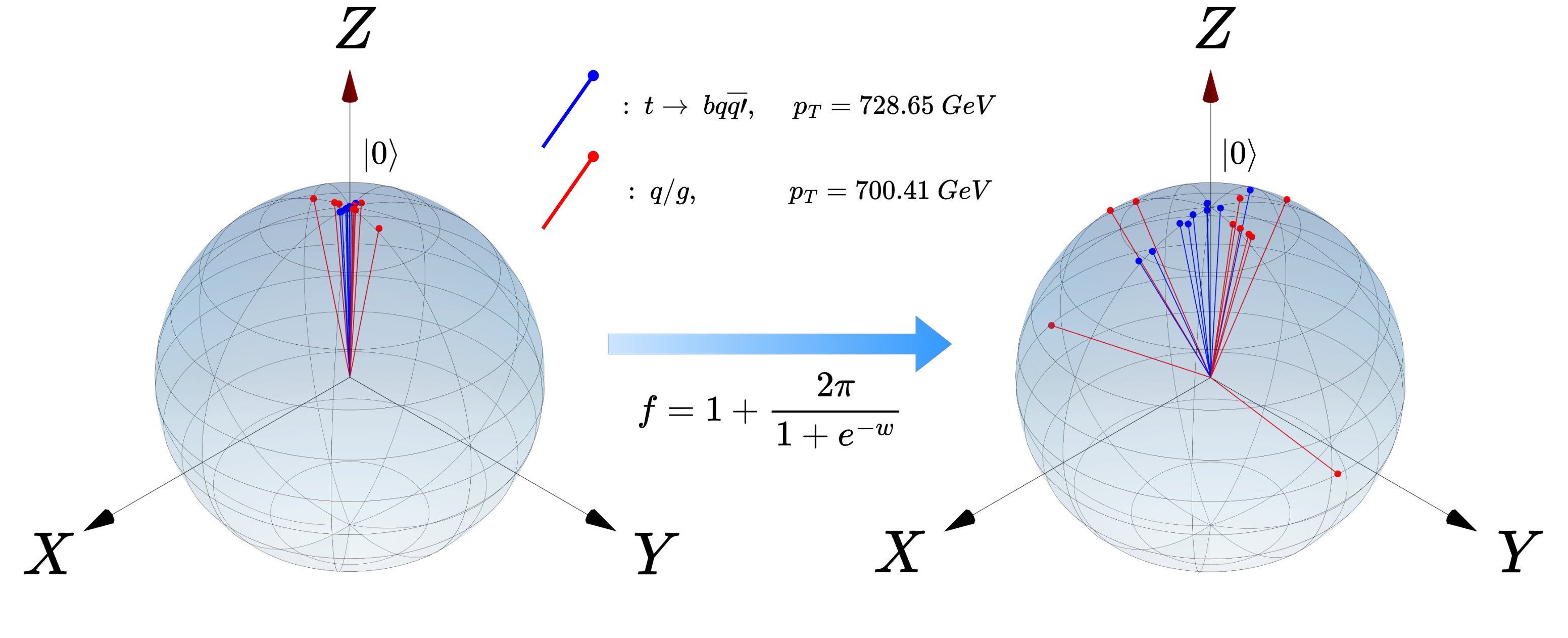}
\caption{Bloch Sphere: Effect of the input scaling described in Eq.~\ref{eq:scale} when applied to the ten hardest particles of a QCD and top jet with comparable jet $p_\mathrm{T}$. The value of $w$ converges to a scaling factor $f=7.268$ for the VQC.}
\label{fig:bloch_inputs}
\end{figure*}

In the above equations, we take the coordinates $\eta$ and $\phi$ relative to the jet axis. While LHC collisions can result in hundreds of final-state particles, we limit our studies to the intrinsic structure of jets rather than entire events due to the limitations of currently available quantum computers and their simulators. Studying the substructure of jets can be a very powerful way to disentangle boosted, hadronically decaying electroweak-scale or beyond-the-standard-model resonances from QCD-induced jets \cite{Marzani:2019hun}. Furthermore, in order to allow for efficient simulation of our quantum circuits on a simulator, we limit our study by using up to ten hardest jet constituents, which are expected to carry most of the information relevant to jet substructure analysis. 
Figure~\ref{fig:bloch_inputs} shows an example of the encoding for the ten hardest particles in a jet from a top quark decay (red) and in a QCD light-flavor jet (blue). Each particle is encoded on a different qubit. As can be seen, the encoding is rooted in the kinematic structure of jets, mapping angular coordinates and transverse momentum to the Bloch sphere to preserve geometric and energy correlations among constituents. The scaling function $f$ in Eq.~\ref{eq:scale} increases the angles on the Bloch sphere.

\paragraph*{\textbf{Anomaly detection and classification with 1P1Q.}}
\label{sec:circuit}
Anomaly detection has emerged as a pivotal tool in model-agnostic searches for new physics, where the only assumption is that  deviations from Standard Model signatures are present~\cite{Belis:2023mqs}. Using classical machine learning, unsupervised learning models based on autoencoders are employed as efficient tools for anomaly detection \cite{ATLAS:2023azi,ATLAS:2023ixc,CMS:anomaly}. These work by compressing the input data into a lower-dimensional latent space and attempting to reconstruct the input, highlighting anomalous events through elevated reconstruction errors.

Quantum autoencoders extend this paradigm to quantum devices~\cite{Ngairangbam:2021yma}. 
A QAE consists of an encoder and a decoder implemented via variational quantum circuits. The encoder compresses the input quantum state into a smaller latent representation, discarding qubits as necessary. The decoder then reconstructs the input from this latent state using the Hermitian conjugate of the encoding operators. To achieve dimensionality reduction in the latent space, QAEs replace a certain number of qubits (hereafter referred to as $N_\text{trash}$ trash states) with $N_\text{ref}=N_\text{trash}$ reference states that are initialized to $\ket{0}$, thus creating an information bottleneck. The unitary transformations within the QAE consist of a combination of parameterized single-qubit rotations and multi-qubit entangling gates, such as the Controlled NOT (CNOT) gate. Using specific examples, QAEs have shown first successes compared to classical AEs regarding training efficiency and performance (e.g., \cite{Ngairangbam:2021yma}). 

\begin{figure*}[ht]
    \centering
    \includegraphics[width=0.497\textwidth]{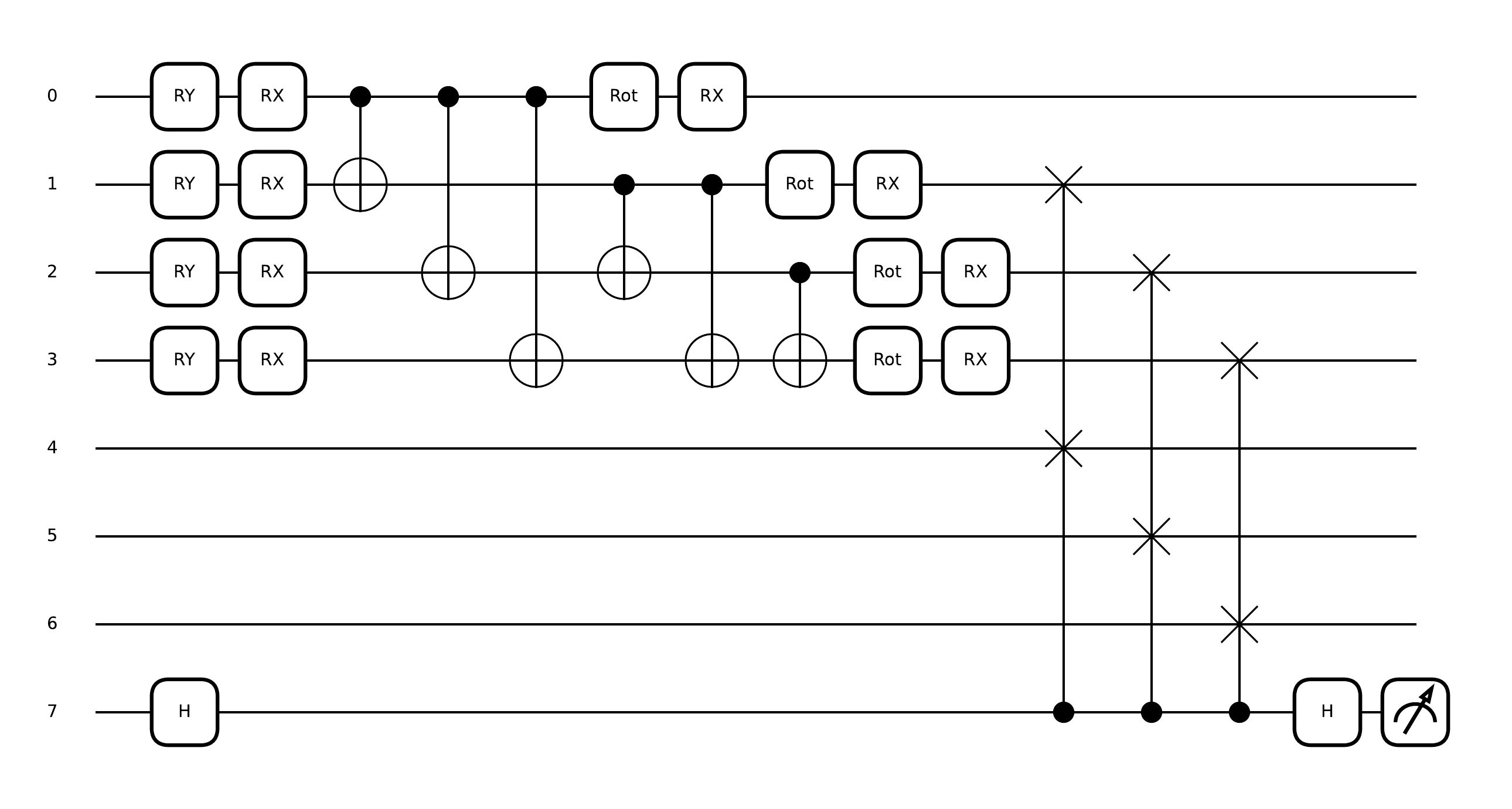}
    \hfill
    \includegraphics[width=0.44\textwidth]{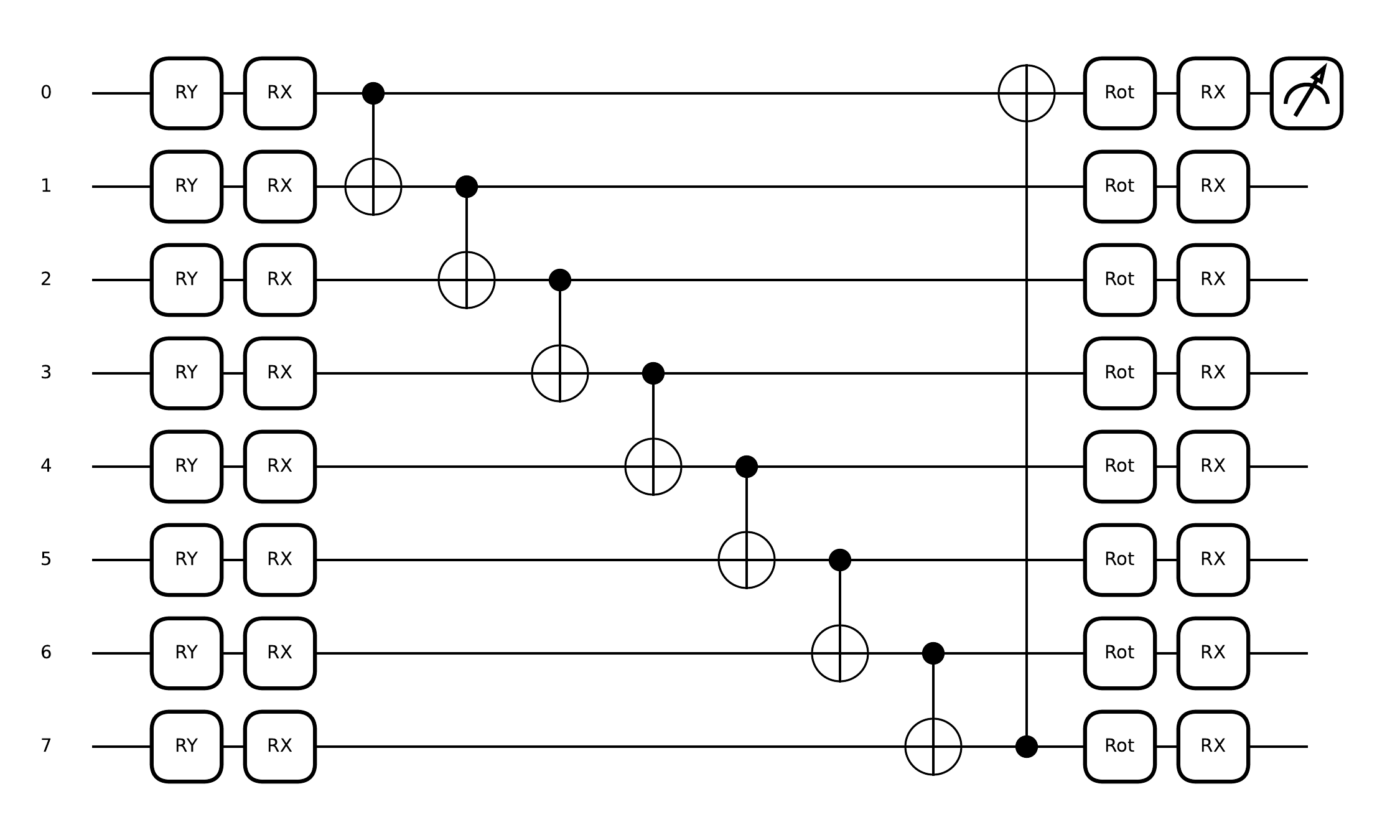}
    \caption{QAE Circuit (left) used for anomaly detection. VQC (right) used for supervised classification. Example circuits with 4 and 8 input particles, respectively.}
    \label{fig:circuits}
\end{figure*}


To demonstrate the applicability of 1P1Q in a QAE, we use Eqs.~\ref{eq:enc1} and \ref{eq:enc2} to encode information into a QAE with an architecture following the left circuit of Fig.~\ref{fig:circuits}. For illustration purposes, the figure shows a $4$-particle-input example, while the QAEs trained are using $6$, $8$, and $10$ particles as input, respectively. To allow the network to learn higher-order non-linear terms, we entangle all possible pairs of qubits using two-qubit CNOT gates. To ensure the network learns an optimal representation of the input space, we apply three parameterized rotations, one along each axis of each qubit. The entanglement and rotation operations are summarized in Eq.~\ref{eq:input}.
\begin{align}
    U(\Theta)=\left(\bigotimes_{i=1}^{N} R_X(\phi_i) R_Y(\theta_i) R_Z(\omega_i)\right) \otimes \left(\bigotimes_{1 \leq i < j \leq N} C_{ij}\right). \label{eq:input}
\end{align}
The QAE is trained to reconstruct jets initiated by light quarks or gluons. We expect this reconstruction quality, defined as the fidelity between the trash and reference states, to differ between signal and background, thus providing a degree of separation.

In addition, we demonstrate that the 1P1Q encoding can just as well be used for classification within a supervised learning paradigm. For this task, we use a VQC \cite{Blance:2020nhl} that learns to separate signal and background classes, in this case jets from hadronically decaying top quarks versus jets initiated by light quarks or gluons. 

The VQC architecture is similar to that of the QAE, with features being first encoded into the circuit using the procedure outlined in Eqs.~\ref{eq:enc1}-\ref{eq:scale}, followed by entanglement operations using two-qubit CNOT gates to allow for the introduction of non-linear terms into the network.  Only adjacent qubits are entangled for the VQC, which we find gives the best performance. Three trainable rotation gates are applied to each qubit, as summarized in Eq.~\ref{eq:VQC_circuit_operation}. Unlike the QAE, which is optimized by maximizing the fidelity between the trash and reference states, the VQC is optimized by performing a measurement on the first qubit which is bound in $[-1,1]$ and comparing it to the ground truth label. The  VQC circuit is shown in Fig.~\ref{fig:circuits} for an input of $N=8$ particles. 

\begin{align}
    U(\Theta) = &\left(\bigotimes_{i=1}^{N} R_X(\phi_i) R_Y(\theta_i) R_Z(\omega_i)\right) \nonumber \\ &\otimes \left(\bigotimes_{i=1}^N C_{i,(i+1)\bmod N}\right)
    \label{eq:VQC_circuit_operation} 
\end{align}


We highlight that for an input space comprising of the $N$ hardest input particles per jet, the QAE requires only $\mathbf{3N+1}$ trainable parameters to learn a suitable reconstruction of the inputs. Likewise, the VQC requires only $\mathbf{3N+2}$ parameters (the extra parameter of the VQC being a trainable bias term that is added to the measurement).

\paragraph*{\textbf{Results and Benchmarking.}}


\begin{figure}[ht]
\includegraphics[width=0.475\textwidth]{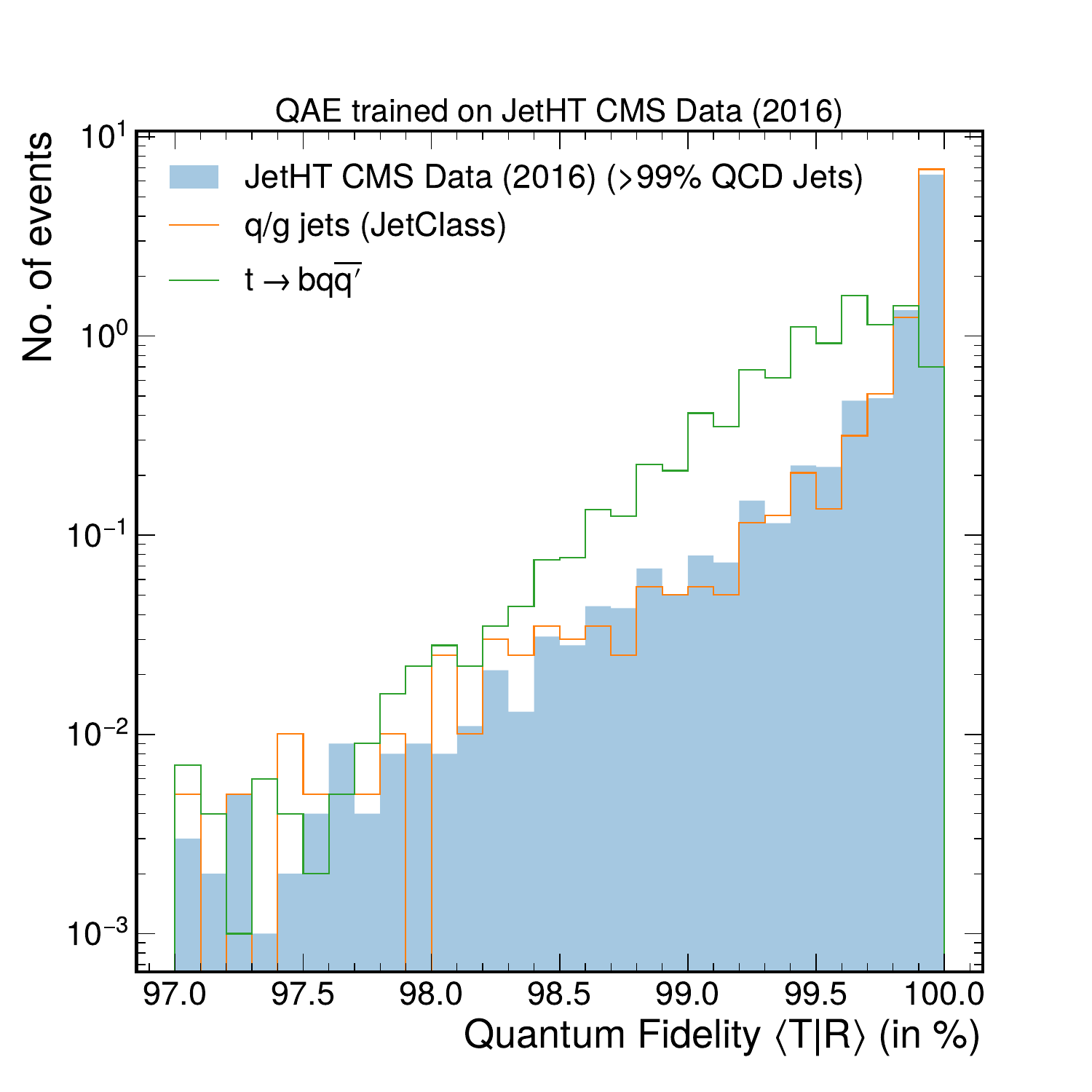}
\caption{Quantum fidelity distributions for a system with $10$ input qubits and a latent space of $2$ qubits.}
\label{fig:aoj_fidelities}
\end{figure}

We train the QAE on two different sets: simulated QCD jets from the JetClass dataset~\cite{JetClass}, and real jet data recorded with the CMS Detector in 2016~\cite{AOJ}, which is $>$99\% pure in QCD jets. For the first time, to our knowledge, Fig.~\ref{fig:aoj_fidelities} shows that a QAE trained on real experimental data is comparable in performance to one trained on simulated events, when applied to the task of differentiating between simulated QCD events and simulated signal models. This suggests that a QAE using 1P1Q encoding, despite not using high-level information, can extract the underlying physics of the jet based on the individual particles encoded onto qubits and is neither corrupted nor biased by the detector response.


\begin{figure*}[ht]
\includegraphics[width=0.975\textwidth]{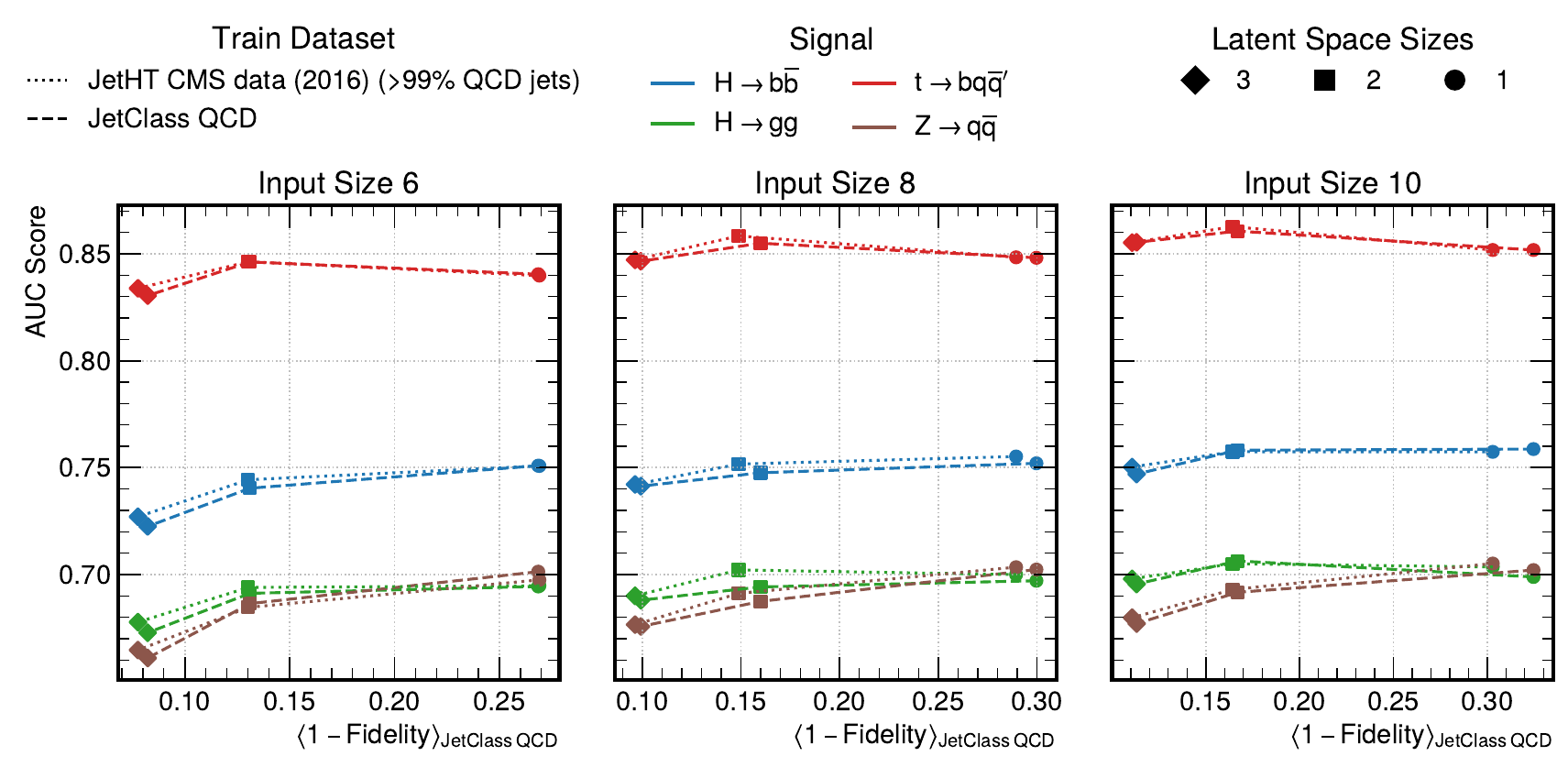}
\caption{AUC scores vs.  $\langle 1 - \text{Fidelity} \rangle_{\text{QCD}}$  for different QAE configurations. Models trained on simulated (dashed) and real CMS data (dotted) show consistent trends, with larger input sizes and higher fidelity loss correlating with improved anomaly detection performance.}
\label{fig:QAEsignal}
\end{figure*}

Figure~\ref{fig:QAEsignal} illustrates the AUC scores as a function of the quantum fidelity metric  $\langle 1 - \text{Fidelity} \rangle_{\text{QCD}}$  for different QAE configurations trained on either simulated JetClass QCD jets (dashed lines) or real CMS data (dotted lines). The three panels correspond to different input sizes, namely 6, 8, and 10 jet constituents, respectively, while different marker shapes indicate varying latent space sizes. Across the various configurations, we observe a trend that higher AUC scores correlate with increased  $\langle 1 - \text{Fidelity} \rangle_{\text{QCD}}$ , indicating that the fidelity loss serves as a useful proxy for anomaly detection performance. The discrimination is particularly strong for top quark jets ($t \to bq\overline{q^{\prime}}$, red), which consistently achieve the highest AUC scores, followed by Higgs decays to bottom quarks ($H \to b\overline{b}$, blue). The similarity in trends between models trained on real and simulated data highlights the robustness of the QAE in learning fundamental jet substructure features independent of dataset origin. Additionally, larger input sizes tend to improve performance, suggesting that incorporating more jet constituents provides richer representations for classification.

To benchmark the performance of the 1P1Q-encoded QAE against a classical counterpart, we trained on simulated QCD jets and considered as anomalies to this background the signals of hadronic $W$ boson, Higgs boson and top quark decays. While the QAE has a simple structure of 10 input qubits, followed by the circuit of Fig.~\ref{fig:circuits} (left) and a latent space of 2, the classical autoencoder (CAE) model is allowed to be significantly larger, consisting of an encoder model containing an input feature vector of size 30, to be able to encode the same number of features encoded on the QAE, followed by dense layers of size $20-16-12$ and a latent space 6. The decoder of the CAE has an identical, yet inverse, structure.

The QAE achieves superior performance compared to the CAE across all signal types. The QAE maintains this advantage despite having only 31 trainable parameters, in contrast to the CAE’s $\sim$2,500 parameters, demonstrating the potential of quantum machine learning to capture relevant physics with a significantly reduced model complexity efficiently.

\begin{table}[t]
\centering
\caption{AUC scores for QAE vs CAE, trained on a 10-particle input space.}
\begin{tabular}{ |c|c|c|c| } 
 \hline
 \diagbox{Algorithm}{Signals} & $W \to q\overline{q^\prime}$ & $H \to b\overline{b}$ & $t \to bq\overline{q'}$ \\ 
 \hline
 QAE & \textbf{0.715} & \textbf{0.774} & \textbf{0.872} \\ 
 \hline
 CAE & 0.671 & 0.739 & 0.858 \\ 
 \hline
\end{tabular}
\label{table:qae_cae}
\end{table}

\begin{figure}[ht]
\includegraphics[width=0.475\textwidth]{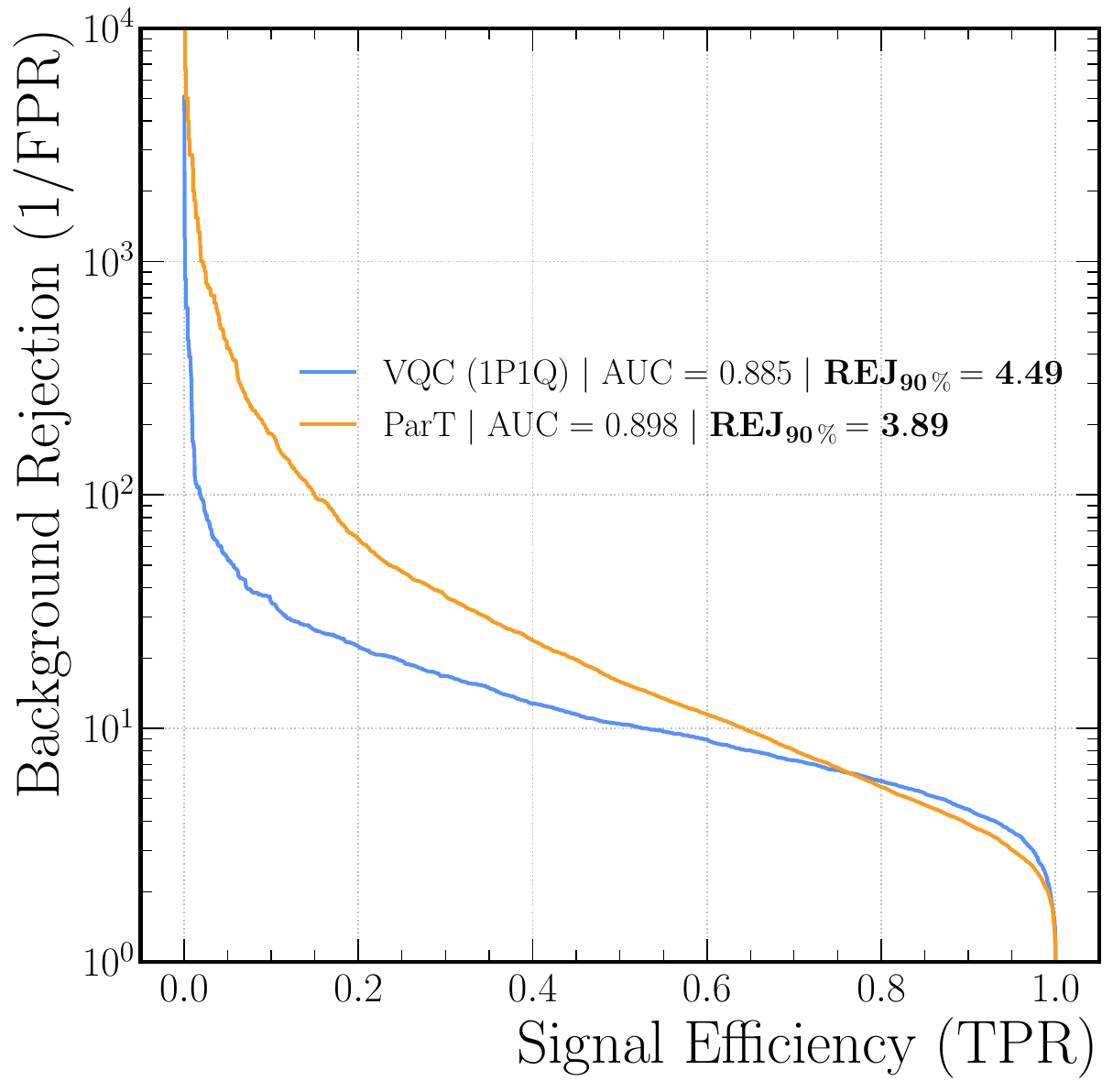}
\caption{Comparison of ROC curves for the VQC and Particle Transformer.
The VQC trained on the 1P1Q-encoded dataset closely matches the performance of the state-of-the-art Particle Transformer, despite using significantly fewer trainable parameters, with both being trained on the same number of events and input size of 10 particles.}
\label{fig:roc_classifier}
\end{figure}

For the supervised classification task, we employ the VQC to distinguish jets originating from top quark decays ($t \to bq\overline{q^{\prime}}$) from those initiated by light quarks or gluons. Figure~\ref{fig:roc_classifier} presents the Receiver Operating Characteristic (ROC) curves comparing the VQC trained on the 1P1Q-encoded dataset against the state-of-the-art Particle Transformer (ParT) classifier~\cite{ParT}, both trained using the 10 hardest particles per jet, and the same dataset of 1000 jets. The ROC curve demonstrates that the VQC achieves strong classification performance, with an AUC score of 0.885. Although the ParT model achieves a slightly higher AUC of 0.898, the VQC remains competitive despite having significantly fewer trainable parameters (32 vs. over 2 million). Furthermore, for high signal efficiencies ($>$80\%), the VQC yields better background rejection than the Particle Transformer. 

This result shows the efficiency and potential of quantum machine learning models in jet classification tasks, especially when using the 1P1Q particle encoding. Such a quantum machine learning method offers competitive performance to classical state-of-the-art frameworks with drastically reduced computational resources.

\paragraph*{\textbf{1P1Q encoding.}}

The expressivity and efficiency of a quantum machine learning approach are jointly defined both by the model (e.g., QAE or VQC) and the encoding step that maps classical data onto a quantum device. Thus, to assess the models' ability to utilize the information provided by the 1P1Q data encoding, we successively reduce the features encoded on each qubit and observe the resulting performance degradation.

\begin{center}
\begin{table}[h]
\centering
\caption{AUC scores vs input features for the VQC and QAE, for the benchmark signal $t\to bq \overline{q^\prime}$.}
\begin{tabular}{ |c|c|c|c|c| } 
 \hline
 \diagbox{Algorithm}{Inputs} &   $(p_\mathrm{T},\eta,\phi)  $ &  $(p_\mathrm{T},\eta)$   & $  (\eta,\phi)$ & $(p_\mathrm{T},\phi)  $ \\ 
 \hline
 VQC & \textbf{0.886} & 0.856 & 0.808 & 0.857 \\ 
 \hline
 QAE & \textbf{0.872} & 0.825 & 0.823 & 0.827 \\ 
 \hline
\end{tabular}
\label{table:qae_vqc}
\end{table}
\end{center}

Table~\ref{table:qae_vqc} shows the impact of different input feature combinations within the 1P1Q encoding scheme on classification performance. The best AUC scores are achieved when all three features ($p_\mathrm{T}, \eta, \phi$) are included, suggesting that the full kinematic variables contribute significantly to the model's performance. When one feature is removed, we observe a performance drop, with the most pronounced reduction occurring when $p_\mathrm{T}$ is excluded. This emphasizes the importance of transverse momentum information in jet classification tasks and suggests that quantum models are particularly effective at utilizing correlations between momentum and angular variables. The robustness of these results across different models further validates the suitability of 1P1Q encoding for high-energy physics applications.

\paragraph*{\textbf{Summary.}}

In this work, we introduced the 1P1Q encoding scheme, a novel approach for representing particle physics data on quantum hardware by assigning each particle to a separate qubit. This encoding allows direct utilization of raw collision event data without classical compression, enhancing the potential expressivity and efficiency of QML models. We demonstrated the effectiveness of this approach using two quantum machine learning models: a QAE for anomaly detection and a VQC for supervised classification. 

Our results show that the QAE successfully differentiates signal jets from background QCD jets, achieving superior performance compared to its classical counterpart while requiring significantly fewer trainable parameters. Furthermore, the VQC exhibits strong classification capability, approaching the performance of state-of-the-art classical models despite its minimal parameter count. By systematically reducing the encoded features, we established that the 1P1Q encoding effectively captures jet substructure information, and performance degradation with reduced feature input underscores the importance of a comprehensive quantum data representation.

For the first time, we validate a quantum machine learning model trained on real experimental data from the CMS detector, demonstrating that quantum approaches can extract meaningful physics insights in a real-world setting. These robust results establish the 1P1Q encoding as a viable and scalable data representation framework for quantum computing applications in particle physics. As quantum hardware continues to advance, the efficiency of the 1P1Q approach is tailored for more intricate and large-scale QML applications in high-energy physics, offering new perspectives for jet classification, anomaly detection, and measurements revealing the quantum mechanical nature of particle production.

\paragraph*{\textbf{Acknowledgements.}}

The authors acknowledge the support of Schmidt Sciences, the Alexander von Humboldt Foundation, and IPPP, as  well as the usage of computing resources on the TOpAS GPU cluster at the Scientific Computing Centre (SCC) KIT, and the RCS Cluster at Imperial College London. The authors also thank the NHR Center Karlsruhe for providing access to computing resources on the high-performance computer HoreKa, which is jointly supported by the German Federal Ministry of Education and Research (BMBF) and the state governments.

\section*{Appendix}
\subsection*{Training setup}

The quantum circuits presented in this study are simulated and optimized using the Quantum Machine Learning library \texttt{pennylane} \cite{Pennylane} with the \texttt{lightning.gpu} and \texttt{lightning.kokkos} devices. 
The trainable circuit parameters are optimized using the classical Adam Optimizer \cite{ADAM}, with a scheduler that periodically decays the learning rate.

To train the QAE, we seek to maximize the fidelity between the output and input states in the subspace relevant for reconstruction: 
\begin{align}
    F &= \sum_{i=1}^{N_\text{ref}} \braket{T_i|R_i} & \forall~~~~ T_i \in H^{T}, R_i \in H^\text{ref} \label{eq:fidelity}
\end{align}
Specifically, following the approach first introduced in \cite{Romero_2017}, we minimize the cost function defined as the negative of the fidelity. This measurement is performed using a SWAP test \cite{Buhrman_2001}, which requires an ancillary qubit initialized to $\ket{0}$. 

In the case of the VQC, we use the expectation value of the Pauli $Z$ observable on the target, which is the first qubit in the circuit. The final VQC prediction (Eq.~\ref{eq:VQC_pred}) is arrived at by adding a classical, trainable bias term $b$ to the expectation value: 

\begin{align}
    f(x) &=\langle q(x)|Z|q(x)\rangle + b \label{eq:VQC_pred} \\
\end{align} 

The circuit optimization is performed using the Mean Squared Error (MSE) between the prediction and the truth label as the loss function.

\subsection*{Datasets}

The \textsc{JetClass} dataset \cite{ParT} is more recent and far larger in scope and size compared to any of its predecessor datasets, and is therefore used to train and benchmark the 1P1Q approach. It contains a total of 100$\mathrm{M}$ jets divided into 10 classes. The background jets are those initiated by light quarks or gluons, while jets arising from decays such as, but not limited to, $t\to b\qqbar$ and $\mathrm{W/Z}\to \qqbar$ are treated as signal jets. The production and subsequent decay of top quarks and the W, Z, and Higgs bosons are simulated with \texttt{Madgraph5}\_\texttt{aMC@NLO} at next-to-leading order precision~\cite{madgraph2014}. Parton showering and hadronisation processes are simulated in \textsc{Pythia8}~\cite{Sjostrand2014}.

To ensure the simulated jets closely resemble those reconstructed by the CMS detector, the detector effects are simulated with \textsc{Delphes}~\cite{delphes2010} using a simplified CMS detector configuration. Jets are clustered utilizing the anti-$k_\mathrm{T}$ algorithm~\cite{cacciari2008antikt} with a distance parameter $R = 0.8$. As additional criteria, only jets with $p_\mathrm{T} \in [500,1000]\,$GeV and pseudorapidity $|\eta| < 2$ are stored. For signal jets, additional quality requirements are imposed to ensure they fully contain the decay products of the initial particles.

The Aspen Open Jets (AOJ) dataset~\cite{AOJ} is derived from the CMS 2016 JetHT datasets~\cite{CMS_Run2016G_JetHT_MINIAOD,CMS_Run2016H_JetHT_MINIAOD} and presented in a structured format specifically optimised for machine learning applications. Although an extensive description of CMS data acquisition and processing falls beyond the scope of this letter, it is important to note that this dataset predominantly comprises jets initiated by light quarks or gluons, with contamination from alternative decay processes such as those involving $\mathrm{W}/\mathrm{Z}$ bosons or top quarks constituting less than $1\%$ of the total sample size. In its entirety, the dataset contains approximately 180$\mathrm{M}$ jets recorded in 2016 with the CMS detector.

For training the QAE, we use a total of 10,000 events for training and 2,500 events for validation. Inference is performed on 10,000 events of background and of each of the following class of events: $H\to b\overline{b}$, $H\to c\overline{c}$, $H\to gg$, $W \to q\overline{q^\prime}$, $Z\to q\overline{q}$ and $t \to bq\overline{q^{\prime}}$. The VQC is trained using 1,000 events and validated on 500 events, with an inference dataset of 10,000 events equally distributed between signal and background classes.

\begin{figure}[ht]
  \centering
  \includegraphics[width=0.47\textwidth]{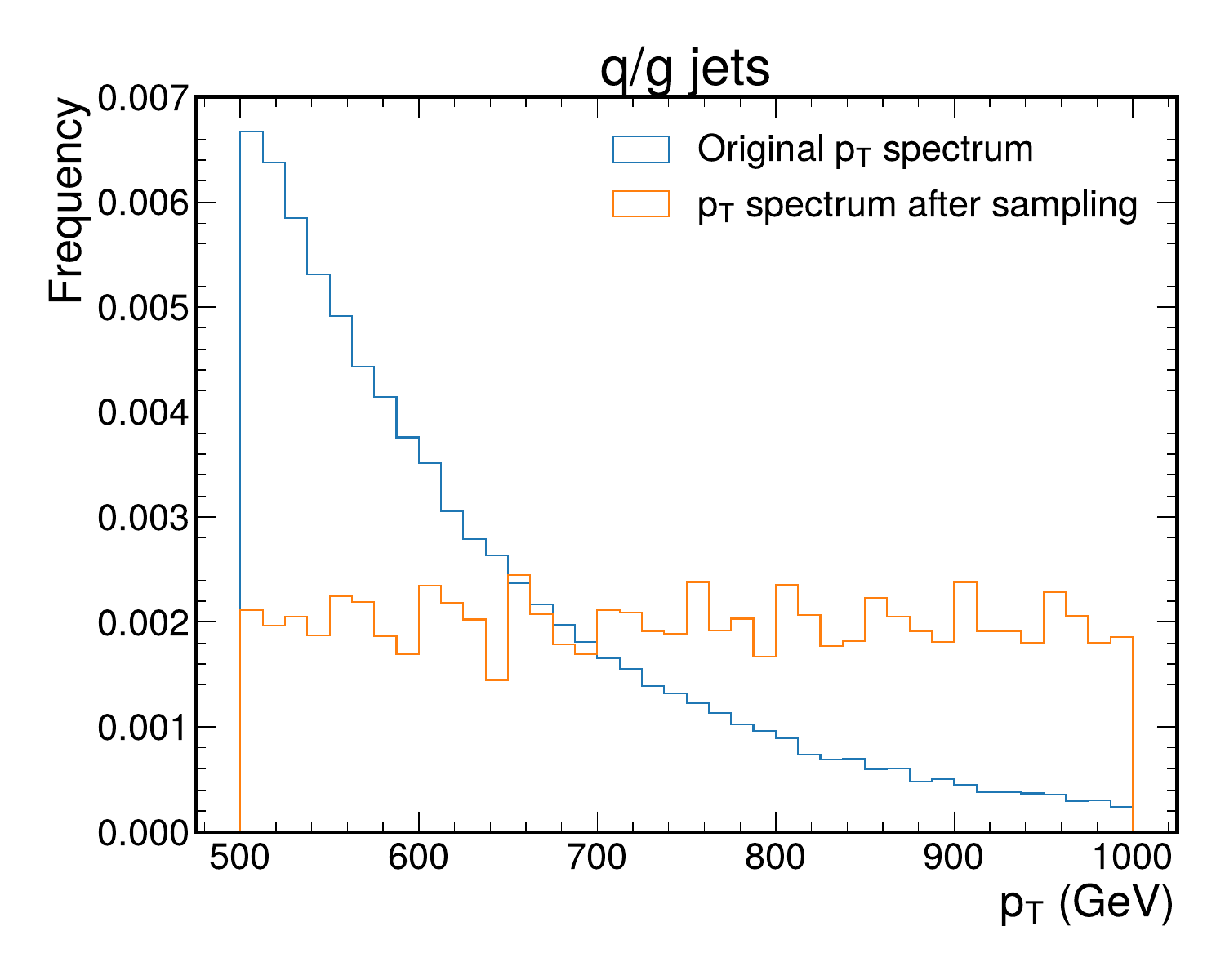}
  \caption{Effect of the \pt~sampling on the input datasets used for training}
  \label{fig:pt_sampling}
\end{figure}

For training the QAE and VQC respectively, we first sample jets such that each class has a flat distribution in jet $p_\mathrm{T}$, in the range $[500,1000]$\,GeV, so as to not bias the training towards the scale of the jet, allowing us to purely focus on jet substructure. The effect of this sampling can be seen in Figure \ref{fig:pt_sampling}. The jet samples used in the inference are also sampled to have the same flat distribution.
\nocite{*}

\bibliography{apssamp}

\end{document}